# Three-axis Hall transducer based on semiconductor microtubes

Alexander Vorob'ev, Anton Chesnitskiy, Alexander Toropov, and Victor Prinz

*Rzhanov Institute of Semiconductor Physics, Russian Academy of Sciences,*
*Acad. Lavrentyev Ave. 13, 630090 Novosibirsk, Russia*

A three-axis Hall transducer based on GaAs/AlGaAs/InGaAs microtubes has been fabricated. The possibility of simultaneous measurement of all the three magnetic field components with such a transducer was demonstrated. It was also shown possible to pack the free-standing parts of the transducer with polydimethylsiloxane for protection of those parts from mechanical actions.

Due to their low cost, small dimensions, and a linear response to magnetic field, Hall transducers find wide application in measurement of moderate and strong magnetic fields (~1 µT and over) and as magnetosensitive components of more complex sensor systems. A simplest Hall transducer in the form of a semiconductor plate provided with four contacts measures the surface-normal component of an external magnetic field. Enhancement of the integration level of electronic devices necessitates, among other things, a corresponding reduction in the dimensions of physical probes and passing over from one-axis sensors, which measure the projection of a quantity to one coordinate axis, to two- and three-axis sensors. In particular, two- and three-axis magnetic field measurements are required in many applications, including: (i) contactless position sensors, and angular and linear movement sensors [1], (ii) evaluation of magnetic field uniformity and magnetic field gradient in weapon detection systems, magnetic traps, etc. [2], (iii) tracking of magnetic markers [3] and guiding of medical tools [4] in medicine and biology, (iv) electronic compasses [5], and (v) scanning Hall probe microscopy [6].

For measuring in-plane components of a magnetic field vector, the so-called vertical Hall devices or integrated magnetic concentrators are used [1]. Vertical Hall devices imply using bulk semiconductors (thickness 10 µm and over), and they do not admit the use of III-V quantum-well heterostructures, which, as a rule, offer a higher sensitivity and better temperature stability. The use of magnetic concentrators increases the dimensions of transducers and brings about into their characteristic some undesired features typical of ferromagnetics such as magnetic saturation and remanent field, leading to response hysteresis and nonlinearity, and also an additional dependence of sensitivity on temperature. To summarize, presently there exists a vital need in a technology that would allow batch fabrication of miniature three-axis Hall sensors, with involved processes being compatible with integration-circuit technology and being free of the limitations outlined above. A reasonable approach here is the assembling of one-axis Hall elements into an integral device fabricated using surface micromachining methods.

It was shown previously that three-dimensional strain-driven micro- and nanostructuring [7] allows the formation of conducting microtubes with several Hall elements having different predefined spatial orientation. Experimentally, the formation of a microtube with two pair of Hall contacts and simultaneous measurement of Hall voltages at two different local values of the surface-normal component of magnetic-field vector have been demonstrated in Ref. 8. Recently, an attempt to fabricate a room-temperature three-axis Hall sensor [9] using the so-called micro-origami version of strain-driven microstructuring [10] was reported. However, three-axis measurements of magnetic field with the help of the fabricated device have not been demonstrated in Ref. 9.

In the present paper, we report on the creation of a three-axis Hall transducer based on GaAs/AlGaAs/InGaAs microtubes, and demonstrate the possibility of measurement, with this transducer, all the three component of magnetic-field vector with the stationary transducer. In addition, we prove it possible to achieve protection of free-standing thin films forming the active part of the transducer from mechanical actions.

In the experiments, a multi-layer strained heterostructure grown by molecular-beam epitaxy on a (001)-oriented GaAs substrate was used. The heterostructure contained, over a GaAs layer, a 10-nm thick AlAs sacrificial layer, a strained 20-nm thick $In_{0.18}Ga_{0.82}As$ layer, and a GaAs quantum well of width 13 nm sandwiched between two $Al_{0.3}Ga_{0.7}As$ layers δ-doped with silicon. Selective etching of the sacrificial layer has allowed us to make the upper film detached from substrate. The profile of the initial heterostructure, and also its band diagram and the distribution of electrons in the film detached from substrate are shown in Fig. 1. For calculations, the 1D Poisson software [11] was used.

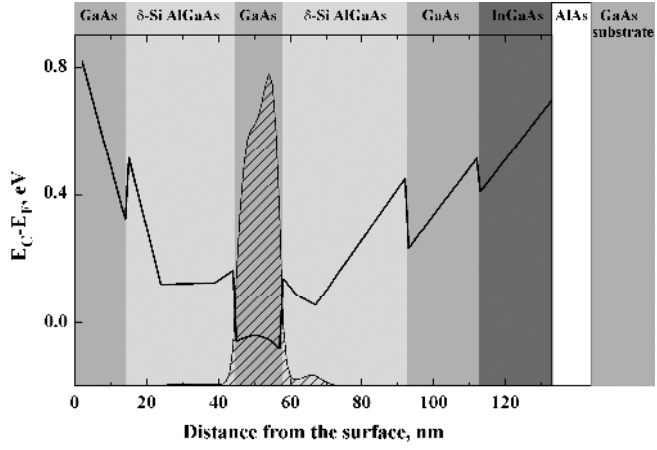

FIG. 1. A diagram illustrating the design of the initial MBE-grown heterostructure. The shades of gray show the layer sequence in the heterostructure, and the thick solid line and the hatched region shows respectively the conduction-band diagram and the distribution of free electrons in the rolled-up film. The redistribution of the strain that occurred when the planar film rolled in a tube was not taken into account.

The three-axis Hall transducer examined in the present study was prepared by directional rolling of a two-level-lithography defined mesastructure [8] as shown in Fig. 2. The fabrication procedure comprised the following steps: (i) definition of two start edges for rolling the necessary tubes by deep etching of the GaAs/AlGaAs/InGaAs film down to the substrate (Fig. 2a); (ii) definition of two Hall bars whose channels were oriented in mutually perpendicular directions [100] and [010], and definition of contact pads by shallow wet etching of the GaAs/AlGaAs film down to the lower Si-doped AlGaAs layer (Fig. 2a); and (iii) release of the strained patterned film by selective etching of the sacrificial layer, resulting in turn in rolling of the film in the preset [100] and [010] directions (Fig. 2b). Due to the anisotropy of the Young modulus in III-V crystals [12], the crystallographic directions <100> are directions preferable for tube rolling since the rolling along them induces less defects in the tube in comparison with other directions. Two-level lithography allows one to implement directional rolling of a film from a starting edge, and also prevents its rolling from the other sides of the mesastructure [13]. The curvature radius of rolled film $R$ is defined by the thicknesses of component layers, and also by structural and mechanical properties of layer materials. In the simplest case of a bilayer film involving two layers of dissimilar materials having thicknesses $d_A$ and $d_B$, lattice constants $a_A$ and $a_B$, Young moduli $E_A$ and $E_B$, and identical Poisson ratios $\nu$ the radius $R$ is given by [14] where $\chi=E_B/E_A$, $\varepsilon=(a_B-a_A)/a_A$.

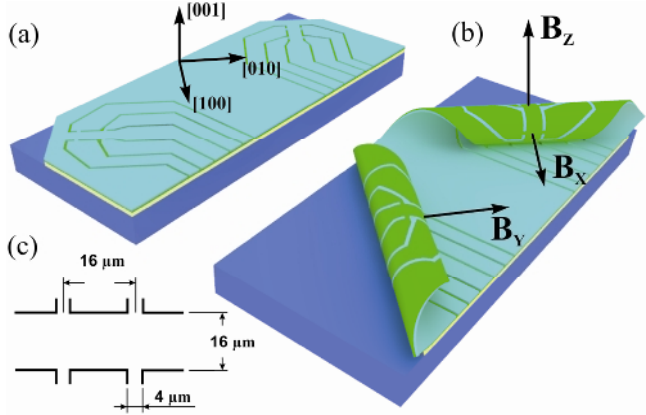

FIG. 2. (a) Lithographically patterned flat Hall bars; (b) Hall bars rolled-up in tubes. The arrows show the orientation of the orthogonal components $B_X$, $B_Y$, and $B_Z$ of the magnetic field vector. (c) Nominal dimensions of the Hall bar.

On detachment from substrate, the lithographically pre-patterned film directionally rolled along the <100> direction in a scroll with visible outer curvature radius *R* of about 10 µm. After rolling, each Hall bar represented a cylindrical sector attached to large-area flat contact pads formed on the substrate (Fig. 2b). The axes of the two formed microtubes were normal to each other. The curvature radius of the microtubes and the dimensions of the lithographic pattern were chosen such that to make the angular spacing between the pairs of Hall contacts in either Hall bar equal to 90°. The etching duration of the sacrificial boundary layer was chosen such that to make all the three Hall junctions lie in orthogonal planes. The width of the channel and the separation between the potential contacts along the channel were 16 µm, and the width of the potential contacts was *l*=4 µm (Fig.2c). Ohmic contacts to the sample were prepared using two alternative methods: (i) sputtering of AuGe/Ni/Au layers onto the flat contact pads and subsequent annealing at T=430° in hydrogen atmosphere given to the sample prior to the formation of microtubes or (ii) application of InSn onto the flat contact pads after the formation of microtubes, followed by subsequent annealing of the sample at T=450° in argon atmosphere. Data measured on samples with contacts of the two types were identical. Images of obtained samples are shown in Fig. 3.

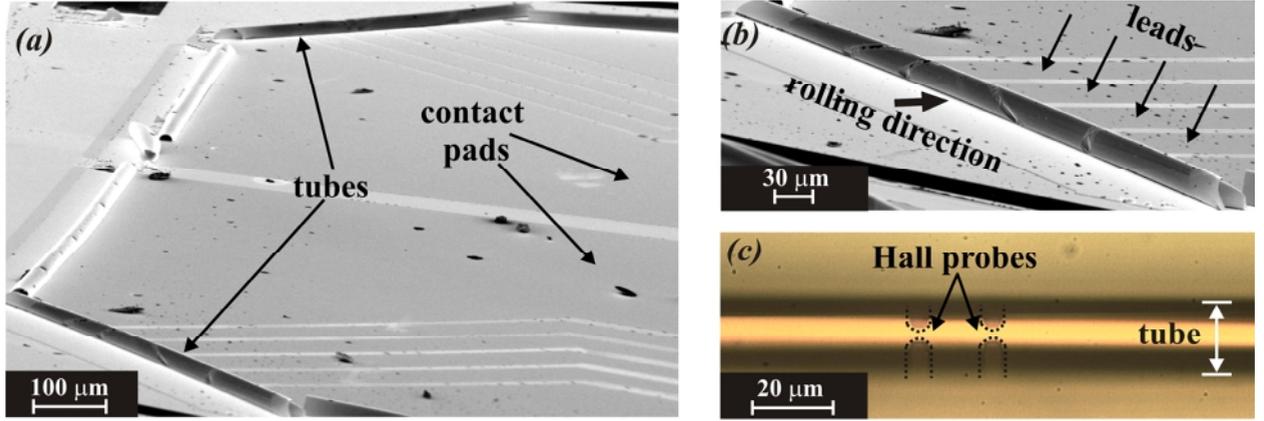

FIG. 3. (Color online) (*a*) SEM image of a sample with two lithographically patterned rolled-up microtubes, the arrows show two microtubes with mutually perpendicular axes, and flat contact pads; (*b*) an enlarged SEM image of the cylindrical shell. The thick arrow shows the rolling direction of the patterned film, and the thin arrows show lithographically defined leads from the rolled-up Hall bar to flat contact pads. Deviations from a perfect cylindrical shape of the tube are due to thinning of the film in etched regions; (c) photo (top view) of the microtube. The arrows show the pair of Hall probes.

Measurements were performed at room temperature in the dark, at direct current *I* up to 150 μA. Measured *I-V* curves for all contact pairs of the samples were linear. For measurement of Hall concentration, the sample was brought to such an orientation that to make the magnetic field directed along the normal to the microtube surface in between one of the Hall-contact pair (see the left inset to Fig. 4). The dependences of the Hall voltage $V_H$ across one of the Hall-contact pairs on the magnetic field for various values of the bias current are shown in the left panel of Fig. 4. It is seen that the dependences are linear throughout the whole examined range of magnetic fields, and their slope varies in proportion to the bias current. The longitudinal resistance of the samples measured in the direction <100> for B=0 was $R_{xx}$=980 Ohm/sq. The found values of the electron sheet density $n_S$, the electron mobility μ along the <100> direction, and the magnetic sensitivity S=($\Delta V_H$/$\Delta$B)/*I* were 1.33x10$^{12}$ cm$^{-2}$, 4.7 x10$^3$ cm$^2$/V s, and 472 Ohm/T, respectively. The absolute magnetic sensitivity $S_A$= $\Delta V_H$/$\Delta$B=*IS* was greater than 0.07 V/T at *I*=150 μA. Strictly speaking, the difference of the Hall-junction shape from a planar one decreases measured Hall voltage $V_H$=<B>*I*/e$n_S$ because the average value of the surface-normal magnetic field vector component <B>=<$B_0$cos  > in the region of the Hall junction is smaller than the strength of the external magnetic field $B_0$ (see the inset in Fig. 4). However, in the examined samples this reduction of Hall resistance $\Delta V_H/V_{H0} = 1 - <\cos\varphi>\big|_{-l/2R}^{l/2R} = 1 - \frac{2R}{l}\sin\frac{l}{2R}$ amounted to less than 0.7%.

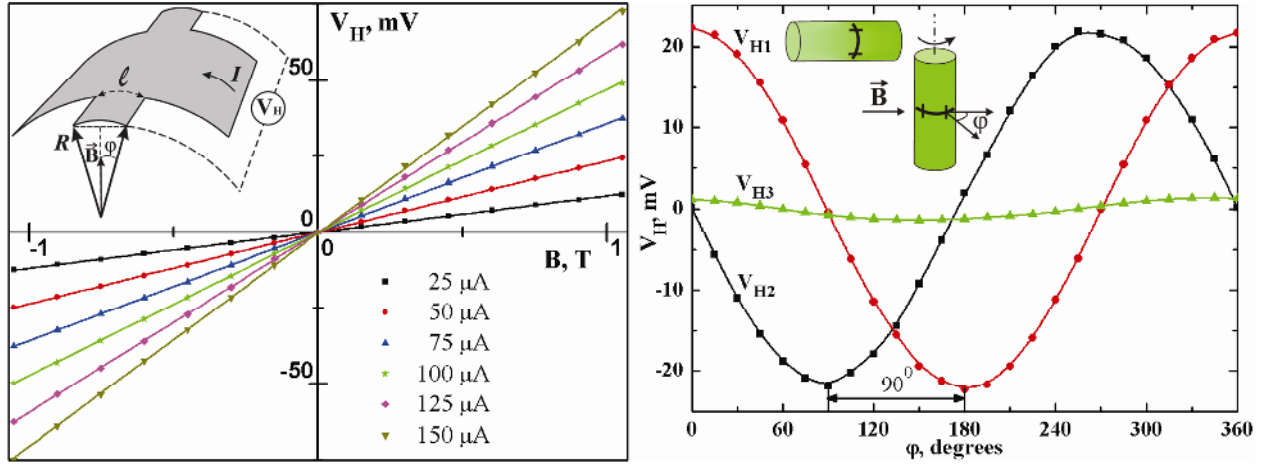

FIG. 4. Left panel: Hall voltage vs magnetic field for different input currents. Right panel: Hall voltages vs rotation angle. The insets show the experimental geometry.

The angular dependences of the Hall voltages measured on the rotation of the sample about one of the coordinate axis are shown in the right panel of Fig. 4. Here, the external magnetic field is directed along the axis Z. The Hall voltages $V_{H1}$, $V_{H2}$, and $V_{H3}$ were measured as a function of the angle of sample rotation $\varphi$ around the axis Y. The curves $V_{H1}(\varphi)$ and $V_{H2}(\varphi)$ are sinusoids having almost identical amplitudes and phase-shifted by 90°, which proves the orthogonal orientation of two cross-shaped Hall junctions on the tube, whose axis is normal to magnetic field direction (see the right inset to Fig. 4). Simultaneously, the Hall voltage $V_{H3}$ is almost independent of the rotation angle $\varphi$, which proves that the external magnetic field vector to a good accuracy lies in the plane of the corresponding cross-shaped Hall junction (an estimate gives for the deflection a value of less than 4°). On rotation of the sample about the two other coordinate axes, similar results with corresponding inter-changes of measured voltages were obtained. Thus, the obtained experimental data demonstrate the possibility of performing three-axis measurements of an external magnetic field with the microtube-based transducer. Small deviations in the orientation of Hall junctions from strictly orthogonal orientation can be compensated for using individual calibration. The accuracy in the orientation of Hall junctions can be improved by, say, lithographic application, onto the film to be rolled, of obstacles that will bring the microtube motion to a stop at a desired place. Of course, the proposed method of assembling one-axis elements into an integral three-axis device is also applicable to transducers other than the Hall ones.

Free-standing thin films are highly sensitive to mechanical actions such as those due to capillary forces exerted by liquid droplets condensing onto the sample. For protection our samples from outside influences, we covered them, with ready microtubes, with polydimethylsiloxane (PDMS), an organosilicon polymer featuring chemical inertness, low surface tension, water-repellency, and transparency. To this end, the sample was immersed into a PDMS solvent and then, into non-polymerized PDMS. The polymer substituted the solvent inside the microtubes, and it also covered the microtubes from outside. For polymerization of PDMS, the sample was given heating. The procedure for mechanical protection of micro- and nano-shells with the help of PDMS will be described in more detail elsewhere [15]. Samples intended for electrical measurements were covered with PDMS after the wiring procedure. The magnetic-field and angular dependences of Hall voltages measured before and after the procedure of sample covering with PDMS were found identical within measurement inaccuracy. Figure 5 shows a photo of a sample covered with PDMS: under the hardened polymer layer about 0.5 mm thick, metallized leads and contact pads, and also undamaged microtubes are seen.

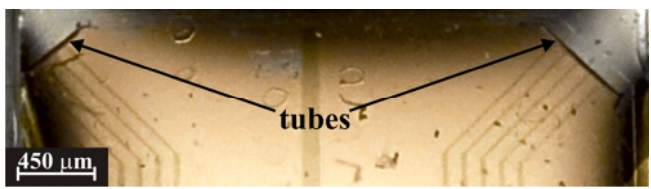

FIG. 5. (Color online) Photo (top view) of a sample covered with PDMS. The arrows show microtubes. The shaded regions have arisen as a result of a reduction of polymer layer thickness at the edges of the sample.

In conclusion, we would like to outline the following potentials offered by the transducer design proposed in the present publication:

1. The proposed method for fabrication of three-axis Hall transducers is fully compatible with the planar technology traditionally used in the production of integrated circuits. As a consequence, it becomes possible to achieve integration of the active element of the transducer within a single chip together with signal-processing circuits, and also to fabricate arrays of identical three-axis sensors that would allow measurement of spatial distributions of magnetic fields and their gradients.

2. The active element of the transducer is scalable, in principle, to sizes of ~100 nm. However, scaling to submicrometer radii will require a corresponding reduction in the thickness of the film to be detached from the substrate. However, the near-surface depletion of GaAs will hardly permit a reduction of the thickness of GaAs-based conducting films below 30 nm with preservation of reasonable carrier concentration and mobility values. In this connection, promising objects for scaling shell-based devices into the nano-scale region involve: (i) shells based on narrow-gap semiconductors [16], (ii) hybrid shells in which semiconductor layers can be used as shape-defining layers, and the active conducting layer can be a metal layer (for instance, a 2-nm thick FePt layer, in which an anomalous Hall effect was observed [17]), and (iii) graphene-based shells [18].

3. Using, instead of III-V microtubes, Si/SiGe microtubes is possible [19, 20]; nowadays, processes for fabrication of Si/SiGe microtubes have received mature development [21].

4. Simplified designs of the transducer described in the present study can also find applications. For instance, one microtube with two Hall junctions can be used as a two-axis transducer, and one microtube with one Hall junction can be used to measure the in-plane magnetic-field component. A tubular shell with a Hall junction seems to be a structure appropriate for monitoring transport of magnetic particles in microfluidic applications.

Acknowledgements

This work was supported by the Russian Foundation for Basic Research (Grants Nos. 12-02-00918 and 12-02-31889).